\documentclass[12pt,preprint]{aastex}

\newcommand{\bdv}[1]{\mbox{\boldmath$#1$}}

\def\au{{\rm AU}} 
 
\def\kms{{\rm km}\,{\rm s}^{-1}}
\def\masyr{{\rm mas}\,{\rm yr}^{-1}}
\def\kpc{{\rm kpc}}

\def\rel{{\rm rel}}

\def\hel{{\rm hel}}
\def\geo{{\rm geo}}
\def\e{{\rm E}}
\def\bpi{{\bdv\pi}}
\def\bmu{{\bdv\mu}}

\def\btheta{{\bdv\theta}}
\def\bphi{{\bdv\phi}}

\def\bv{{\bf v}}

\begin{document}
\title{First Space-based Microlens Parallax Measurement of an Isolated Star:
{\it Spitzer} Observations of OGLE-2014-BLG-0939}

\author
{J.~C.~Yee$^{1}$,
A.~Udalski$^{2}$,
S.~Calchi Novati$^{3,4,5}$,
A.~Gould$^{6}$, 
S.~Carey$^{7}$,
R.~Poleski$^{2,6}$,
B.S.~Gaudi$^{6}$,
R.W.~Pogge$^{6}$
J.~Skowron$^{2}$,
S.~Koz{\l}owski$^{2}$,
P.~Mr{\'o}z$^{2}$,
P.~Pietrukowicz$^{2}$,
G. Pietrzy{\'n}ski$^{2,8}$
M.K.~Szyma{\'n}ski$^{2}$,
I.~Soszy{\'n}ski$^{2}$,
K.~Ulaczyk$^{2}$,
{\L}.~Wyrzykowski$^{2,9}$\\
\normalsize{$^{1}$Harvard-Smithsonian Center for Astrophysics, 60 Garden St., Cambridge, MA 02138, USA} \\
\normalsize{$^{2}$Warsaw University Observatory, Al.~Ujazdowskie~4, 00-478~Warszawa, Poland} \\ 
\normalsize{$^{3}$NASA Exoplanet Science Institute, MS 100-22, California Institute of Technology, Pasadena, CA 91125, USA\footnote{Sagan visiting fellow}} \\
\normalsize{$^{4}$Dipartimento di Fisica ``E. R. Caianiello'', Universit\`a di Salerno, Via Giovanni Paolo II, 84084 Fisciano (SA),\ Italy} \\
\normalsize{$^{5}$Istituto Internazionale per gli Alti Studi Scientifici (IIASS), Via G. Pellegrino 19, 84019 Vietri Sul Mare (SA), Italy}
\normalsize{$^{6}$Department of Astronomy, Ohio State University, 140 W. 18th Ave., Columbus, OH  43210, USA} \\
\normalsize{$^{7}$Spitzer Science Center, MS 220-6, California Institute of Technology,Pasadena, CA, USA} \\
\normalsize{$^{8}$Universidad de Concepci{\'o}n, Departamento de Astronomia, Casilla 160--C, Concepci{\'o}n, Chile} \\
\normalsize{$^{9}$Institute of Astronomy, University of Cambridge, Madingley Road, Cambridge CB3 0HA, UK} \\
}


\begin{abstract}
We present the first space-based microlens parallax measurement
of an isolated star.  From the striking differences in the lightcurve
as seen from Earth and from {\it Spitzer} ($\sim 1\,\au$ to the West),
we infer a projected velocity $\tilde v_\hel\sim 250\,\kms$, which
strongly favors a lens in the Galactic Disk with mass 
$M= 0.23\pm 0.07\,M_\odot$ and distance $D_L= 3.1\pm 0.4\,\kpc$.  
An ensemble of
such measurements drawn from our ongoing program could be used to 
measure the single-lens mass function
including dark objects, and also is necessary for measuring
the Galactic distribution of planets since the ensemble reflects
the underlying Galactic distribution of microlenses.  We study
the application of the many ideas to break the four-fold degeneracy
first predicted by Refsdal 50 years ago.  We find that this degeneracy
is clearly broken, but by two unanticipated mechanisms.

\end{abstract}


\section{{Introduction}
\label{sec:intro}}

When modern microlensing experiments were proposed toward the Large
Magellanic Cloud \citep{pac86} and the Galactic Bulge \citep{pac91,griest91},
it was believed that the only information that could be extracted about 
the lens mass $M$, distance $D_L$, and transverse motion $\bmu_\geo$
would come through
their combination in a single measured parameter, the Einstein timescale,
\begin{equation}
t_\e = {\theta_\e\over\mu_\geo};
\qquad
\theta_\e^2\equiv \kappa M \pi_\rel;
\qquad
\kappa\equiv {4 G\over c^2\au}\simeq 8.14{{\rm mas}\over M_\odot}.
\label{eqn:tedef}
\end{equation}
Here $\theta_\e$ is the angular Einstein radius,
$\pi_\rel = \au(D_L^{-1}-D_S^{-1})$ is the lens-source relative parallax, and
$\mu_\geo$ is the lens-source relative proper motion in the Earth frame
at the peak of the event.  This would imply, in particular, that 
individual masses could be estimated only to within an order of magnitude
(e.g., Figure~1 of \citealt{gould00}).  

It was quickly realized, however, that if two additional
potentially observable quantities could be measured, $\theta_\e$ and
the ``microlens parallax vector'' $\bpi_\e$, then these three quantities
could be disentangled \citep{gould92},

\begin{equation}
M = {\theta_\e\over\kappa \pi_\e};
\qquad
\pi_\rel = \pi_\e\theta_\e;
\qquad
\bmu_\geo = {\theta_\e\over t_\e}{\bpi_{\e,\geo}\over\pi_\e}.
\label{eqn:meqn}
\end{equation}
In modern notation, the microlens parallax vector is given by \citep{gould00b},
\begin{equation}
\bpi_\e \equiv {\pi_\rel\over \theta_\e}\,{\bmu\over\mu}.
\label{eqn:piedef}
\end{equation}
Its amplitude quantifies the lens-source relative displacement in the 
Einstein ring due to motion of the observer, while its direction specifies
the orientation of this displacement as the event evolves.  Hence,
$\bpi_\e$ is in principle measurable from photometric deviations of
the event relative to what is expected from rectilinear motion. See
Figure 1 of \citet{gouldhorne} for a didactic explanation.  

While both $\theta_\e$ and $\bpi_\e$ are important, measurements
of $\bpi_\e$ are more pressing for the following three reasons.
First, $\theta_\e$ is very frequently measured 
``automatically'' in planetary and binary events.  Hence, 
$\pi_\e$ is the crucial missing link to obtain individual masses for these
high priority events, i.e.,  those for which individual
masses are the most important.  Second, $\theta_\e$ is very rarely 
measurable in single-lens events, which means that measuring $\pi_\e$
is the best way to obtain strong
statistical constraints on masses of the much larger population of 
(dark and luminous) single lenses.  Third, while $\pi_\e$ and 
$\theta_\e$ appear symmetrically
in Equation~(\ref{eqn:meqn}), $\pi_\e$ is actually much richer
in information than $\theta_\e$.  This is because the great majority
of lenses observed toward the Galactic Bulge have similar proper motions
within a factor $\sim 2$ of $\mu\sim 4\,\masyr$.  Thus, in the limit
that all microlens proper motions had exactly this
value, a measurement of $\theta_\e = \mu t_\e$ would contain
no additional information, while $\pi_\e$ would completely determine the mass
$M=\mu t_\e/\kappa \pi_\e$.  Although this limit does not strictly apply,
an ensemble of $\pi_\e$ measurements would constrain the mass function
very well \citep{han95}.

There are two broad classes of methods by which parallax might be
measured. The first is to make a single time series from an accelerated 
platform, either Earth \citep{gould92,alcock95,poindexter05}, 
or a satellite in low-Earth 
\citep{honma99} or geosynchronous \citep{gould13} orbit.  The second
is to make simultaneous observations from two (or more) observatories,
either on two platforms in solar orbit \citep{refsdal66}, or located
at several places on Earth \citep{hardy95,holz96,gould97}.
However, with one exception, all of these methods are either subject to 
extremely heavy selection bias or are impractical for the present and near
future.  In particular, out of more than 10,000 microlensing events
discovered to date, fewer than 100 have $\pi_\e$ measurements derived from
Earth's orbital motion, and these are overwhelmingly events due to
nearby lenses and with abnormally long timescales (e.g., 
Table~\ref{tab:ulens_1} of 
\citealt{gould10}).  Only two events have terrestrial parallax
measurements \citep{ob07224,ob08279},
and \citet{gouldyee13} showed that these are subject to
even more severe selection so that even the two recorded measurements
is unexpectedly high.

Hence, the only near-term prospect for obtaining a statistical sample of
microlens parallaxes from which to derive an unbiased mass function,
as originally outlined by \citet{han95}, is by combining Earth-based
observations with those of a satellite in solar orbit.   There are
several major benefits to such a study.  First, it is the only
way to obtain a mass-based census of stellar, remnant, and planetary
populations.  Several components of this population are dark or essentially
dark including free-floating planets, brown dwarfs, neutron stars, and
black holes and therefore are essentially undetectable by any other
method unless they are orbiting other objects.  In addition, even the
luminous-star mass function of distant populations (e.g., in the 
Galactic Bulge) is substantially more difficult to study photometrically
than is generally imagined.  For example, a large fraction of stars
are fainter components in binary systems, with separations that are
too small to be separately resolved, but whose periods are too long
(or primaries too faint) for study by the radial velocity technique.

In 2014,
we were granted Director's Discretionary Time for
a 100 hr pilot program to determine
the feasibility of using {\it Spitzer} as such a parallax satellite
for microlenses observed toward the Galactic Bulge.  The main objective
of this program was to measure lens masses in planetary events.
However, especially in view of the fact that there is generally no
way to distinguish such planetary events from single-lens events
in advance, a secondary goal was to obtain parallaxes for
an ensemble of single-lens events.  Prior to this program, there
had been only one space-based parallax measurement, which was for
a binary lens toward the Small Magellanic Cloud \citep{dong07}.

Here we report on the first space-based parallax measurement of an isolated
lens, OGLE-2014-BLG-0939L.  This measurement serves as a pathfinder
and as a benchmark to test ideas that have been discussed
in the literature for almost 50 years about how to resolve degeneracies
in such events.

{\subsection{Degeneracies in Space-Based Microlens Parallaxes}
\label{sec:degen}}

As already pointed out by \citet{refsdal66}, space-based microlensing
parallaxes are subject to a four-fold discrete degeneracy.  This is
because, to zeroth order, the satellite has a fixed separation from
Earth projected on the plane of the sky ${\bf D}_\perp$, and hence they
measure identical Einstein timescales 
$t_\e=t_{\e,\rm sat}=t_{\e,\oplus}$.  Since the flux evolution $F(t)$
of a single-lens microlensing event is given by \citep{pac86},
\begin{equation}
F(t) = F_S A(t) + F_B;
\qquad A[u(t)] = {u^2+2\over\sqrt{u^4 + 4u^2}};
\qquad
[u(t)]^2 = u_0^2 + {(t-t_0)^2\over t_\e^2},
\label{eqn:pac}
\end{equation}
they are therefore distinguished only by different times of peak $t_0$
and different impact parameters $u_0$ (in addition to the
nuisance parameters $F_S$ and $F_B$, the source and blended-light fluxes,
respectively).
The microlens parallax $\bpi_\e$ can nominally be derived from these
differences,
\begin{equation}
\bpi_\e = {\au\over D_\perp}\biggl({\Delta t_0\over t_\e},\Delta u_0\biggr),
\label{eqn:sbpar}
\end{equation}
where $\Delta t_0=t_{0,\rm sat}-t_{0,\oplus}$, $\Delta u_0=u_{0,\rm sat}-u_{0,\oplus}$,
and where the $x$-axis of the coordinate system is set by the
Earth-satellite vector ${\bf D}_\perp$.
The problem is that while $\Delta t_0$ 
is unambiguously determined from this procedure,
$u_0$ is actually a signed quantity whose amplitude is recovered from
simple point-lens events but whose sign is not (since it appears only 
quadratically in Equation~(\ref{eqn:pac})).  Hence, there are two
solutions $\Delta u_{0,-,\pm} = \pm(|u_{0,\rm sat}|-|u_{0,\oplus})|)$ for which
the satellite and Earth observe the source trajectory on the
same side of the lens as each other (with the ``$\pm$'' designating which
side this is), and two others  
$\Delta u_{0,+,\pm} = \pm(|u_{0,\rm sat}|+|u_{0,\oplus})|)$ for which the source
trajectories are seen on opposite sides of the lens 
(\citealt{gould94}, Figure~1).

For most applications, only the second of these two degeneracies
is important.  That is, the two solutions $\Delta u_{0,-,\pm}$ have
the same amplitude of parallax $\pi_\e$ (as do the two solutions
$\Delta u_{0,+,\pm}$) and so yield the same lens mass and distance.
In each case, the solutions differ only in the direction of lens-source
motion, which is usually not of major interest.  However, the two
sets of solutions can yield radically different $\pi_\e$.  Hence,
if these sets of solutions really cannot be distinguished, the
value of the parallax measurement is seriously undermined.  As a result,
considerable work has been applied over two decades to figuring out
how to break these degeneracies.

Before reviewing this work, however, one should note an important exacerbation
of the underlying problem.  If the four solutions are placed in the
$(\Delta t_0/t_\e,\Delta u_0)$ plane, they of course all lie along a vertical
line of constant $\Delta t_0$.  As pointed out by \citet{gould95},
the error ellipses are also elongated in the vertical direction.  This
is because $u_0$ is strongly correlated with the nuisance parameters
$F_S$ and $F_B$ (since all three enter Equation~(\ref{eqn:pac})
symmetrically in $(t-t_0)$) while $t_0$, which enters anti-symmetrically,
is not strongly correlated with other parameters.  This continuous degeneracy
enhances the probability that the discretely degenerate solutions
will overlap and become a continuous degeneracy.

Four ideas have been proposed to break the $\Delta u_0$ four-fold degeneracy.

{\subsubsection{Measurement of $\Delta t_\e$}
\label{sec:deltate}}
\citet{gould95} proposed to break the degeneracy by using the fact that
the Earth-satellite separation changes with time, and therefore
$t_{\e,\rm sat}\neq t_{\e,\oplus}$.  For near-circular,
near-ecliptic orbits (characteristic of both {\it Spitzer} and {\it Kepler}),
this works quite well for targets near the ecliptic poles 
\citep{boutreux96} because the difference in timescales $\Delta t_\e$
is directly proportional to $\Delta u_0$.  However, it becomes increasing
problematic for targets close to the ecliptic, like the Galactic
Bulge \citep{gaudi97}, because for targets directly on the ecliptic,
$\Delta t_\e$ does not depend at all on $\Delta u_0$ to linear order.
That is, $\Delta u_0$ completely disappears from Equation (2.3) of
\citet{gould95}.

{\subsubsection{Photometric Alignment of Space and Ground Observations}
\label{sec:photo}}
\citet{gould95} also proposed to equip the satellite with a camera
having identical photometric response to one on the ground, which
would guarantee that $F_{S,\rm sat}=F_{S,\oplus}$ and so effectively
insulate $\Delta u_{0,-,\pm}$ from uncertainties in $F_S$ by forcing
the two $\Delta u_{0,-,\pm}$ solutions to move together in a highly-correlated way
as $F_S$ is varied over its allowed range.  While this idea would be quite
difficult to implement, \citet{mb11293} demonstrated that observations
in different bands could be aligned quite tightly with each other
based on color-color diagrams of reference stars.  As a practical
matter, it is not obvious that this technique can be applied to {\it Spitzer}
observations because \citet{mb11293} predicted $F_S$ for a certain
band by {\it interpolating} between two other measured bands, whereas
predicting {\it Spitzer}'s $3.6\,\mu$m $F_S$ requires considerable
{\it extrapolation} from ground-based bands.

{\subsubsection{Combining 1-D Parallaxes from Space and Ground}
\label{sec:1Ds}}
\citet{gould99} suggested that the robust 
one-dimensional (1-D) parallax information along the ${\bf D}_\perp$ 
(i.e., $\Delta t_0$) direction from Earth-satellite observations 
could be combined with robust 1-D information along the direction of
Earth's projected acceleration from ground-based observations
\citep{gmb94} to break the $\Delta u_0$ degeneracy.  This idea was
specifically motivated by the possibility of {\it Spitzer} parallax 
observations toward the Magellanic Clouds, which are at high ecliptic 
latitude  where these two directions are nearly orthogonal.
As he noted, it is substantially more difficult to apply this approach
toward the Bulge where the two directions are close to parallel.   

{\subsubsection{High-Magnification Events (As Seen From Earth)}
\label{sec:himag}}
\citet{gouldyee12} pointed out that for sufficiently high-magnification events
as observed from Earth $(|u_{0,\oplus}|\ll 1)$, we have
$|u_{0,\oplus}|\ll |u_{0,\rm sat}|$ and therefore 
$|\Delta u_{0,-,\pm}|\simeq |\Delta u_{0,+,\pm}|$, so that there is no degeneracy
in the amplitude of $\bpi_\e$, although the direction degeneracy persists.
Moreover, if one of the satellite observations were actually made near
$t_{0,\oplus}$, then only 1--3 satellite observations would be
required.  They therefore
advocated targeting such events.  However, since OGLE-2014-BLG-0939
was not a high-magnification event, this idea is not directly relevant
here and is included only for completeness.

Because this is the first space-based parallax measurement for a
single-lens event, we systematically study the role of all these
ideas (except the last) for both characterizing and breaking the degeneracies
in practice.  We note at the outset that two of these methods are
adversely affected by the Bulge being close to the 
ecliptic, and that this problem
is more pronounced for OGLE-2014-BLG-0939 than for typical events
because it lies just $+2.0^\circ$ from the ecliptic, i.e., about
3 times closer to it than Baade's Window.

{\section{Observations}
\label{sec:obs}}

{\subsection{OGLE Observations}
\label{sec:ogleobs}}

On 2014 May 28, the Optical Gravitational Lens Experiment (OGLE)
alerted the community to a new microlensing event OGLE-2014-BLG-0939
based on observations with the 1.4 deg$^2$ camera on its 
1.3m Warsaw Telescope at the Las Campanas Observatory in Chile
using its Early Warning System (EWS) real-time event detection
software \citep{ews1,ews2}.  Most observations were in $I$ band, but with
three $V$ band observations during the magnified portion of the event
to determine the source
color. At equatorial coordinates (17:47:12.25, $-21$:22:58.7), this
event lies in OGLE field BLG630, which implies that it is observed
at relatively low cadence, roughly once per two nights.

\subsection{Spitzer Observations}

The structure of our {\it Spitzer} observing protocol is described
in detail in Section 3.1 of \citet{ob140124}.  In brief, observations were made
during 38 2.63 hr windows between HJD$^\prime\equiv$HJD$-245000 = 6814.0$
and 6850.0.  Each observation consisted of six dithered 30s exposures
in a fixed pattern using the $3.6\,\mu$m channel on the IRAC camera.  
Observation
sequences were upload to {\it Spitzer} operations on Mondays at UT 15:00,
for observations to be carried out Thursday to Wednesday (with slight
variations).  As described in \citet{ob140124}, JCY and AG balanced various
criteria to determine which targets to observe and how often.  In general,
there were too many targets to be able to observe all viable targets 
during each epoch.

At the decision time (June 2 UT 15:00, HJD$^\prime$ 6811.1) for the
first week of {\it Spitzer} observations, OGLE-2014-BLG-0939 was
poorly understood, with acceptable fits having Earth-based peaks over the range
$6807\la t_{0,\oplus} \la 6845$, i.e., 
from well before to (effectively, see below)
the end of
the {\it Spitzer} observing interval.  Nevertheless, it was put in
the ``daily'' category and observed during all eight epochs, in part because
the source was bright, implying good precision {\it Spitzer} photometry. 
The following week, it was degraded to ``low'' priority
because it was unclear that it would have low enough $u_0$ for an
effective parallax measurement, and if $u_0$ were low enough, 
the peak would be well in
the future.  However, due to a transcription error, it
was left in the ``daily'' file and observed during all six epochs.
By upload time for the third week it was clear first that
$t_{0,\oplus}$ would occur during or near these observations and second
that the amplitude would be low (i.e., relatively high impact parameter
$u_{0,\oplus}\sim 1$).  These considerations
pulled in opposite directions, resulting in
``moderate'' priority and so observations during
six out of eight epochs.  The fact that the predicted peak (from Earth)
was expected to occur at the beginning of the fourth week led to classifying
the event as ``daily'', and so it was observed in all seven epochs.
Because OGLE-2014-BLG-0939 lies relatively far to the West, it moved
out of the {\it Spitzer} observing window (set by the Sun-angle) during
the final week.  Hence it was observed during all four of the available
epochs (out of eight total).  Hence, OGLE-2014-BLG-0939 was observed
relatively uniformly, close to once per day, during the entire
interval that it was observable, from 6814.1 to 6845.7.

{\section{Lightcurve Analysis}
\label{sec:anal}}

The analysis of the lightcurve is straightforward because
the magnification for a single-lens can be written in closed form
(Equation~(\ref{eqn:pac})), i.e., $A=(u^2+2)/(u^4 + 4u^2)^{1/2}$.
While the argument $u$ in this equation is not as simple as in the
case of rectilinear motion illustrated in Equation~(\ref{eqn:pac}),
the deviations from that formula due to Earth's motion are easily
incorporated \citep{gould04}.  {\it Spitzer}'s offset from the
center of Earth is treated just as any other observatory, except
that it is much larger, i.e., of order AU rather than $R_\oplus$.
We adopt the inertial frame that is coincident with the position and velocity
of Earth at the peak of the event, i.e., ${\rm HJD}^\prime=6836.06$.
Any frame will yield equivalent results (after a suitable transformation
of parameters).  However, this (quite standard) geocentric 
frame permits direct comparison
with the results from Earth-only observations, which turns out to be
crucial to understanding the degeneracies.

\begin{figure}
\plotone{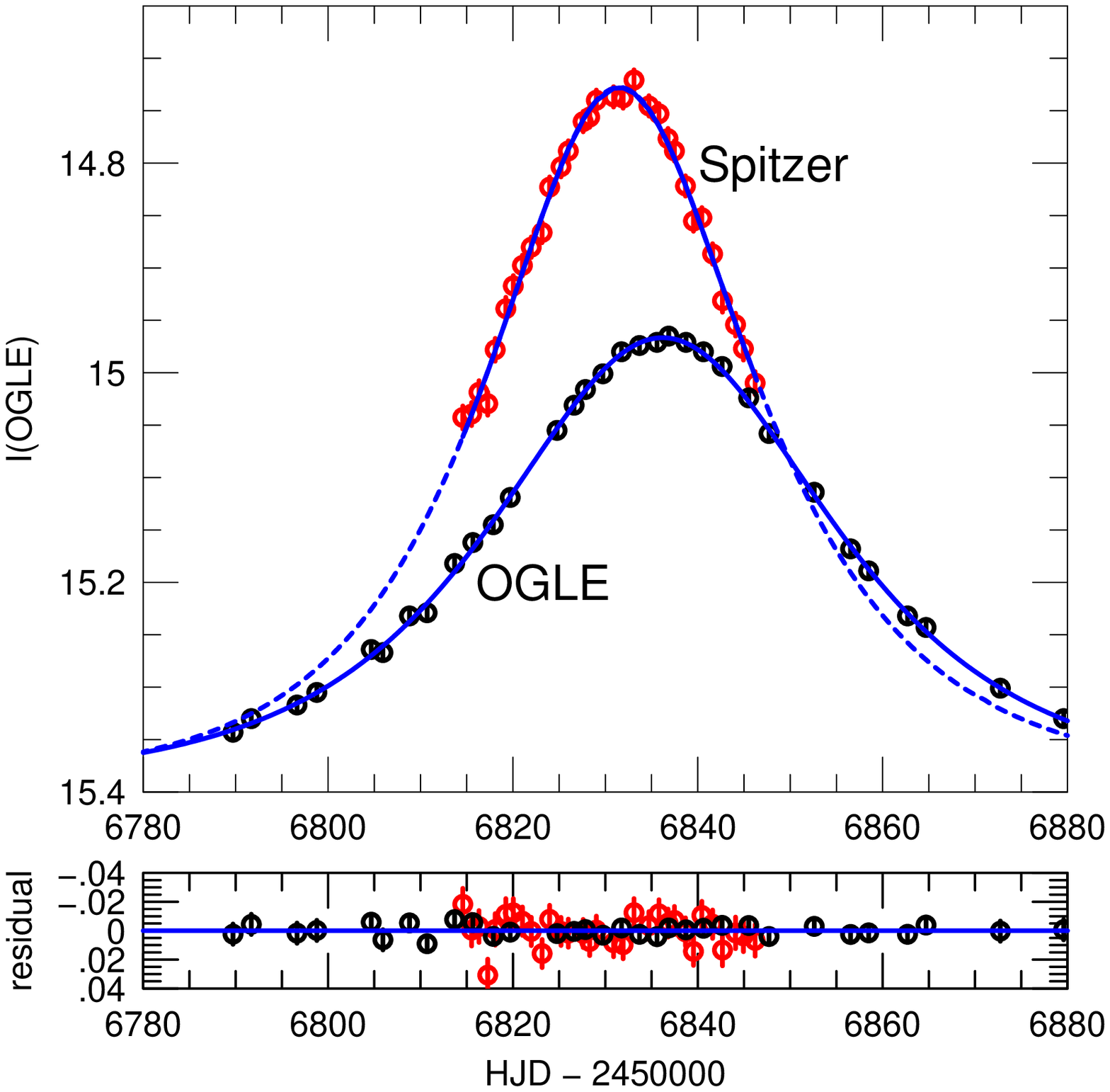}
\caption{Lightcurve of OGLE-2014-BLG-0939 as seen by OGLE from Earth (black)
and {\it Spitzer} (red) $\sim 1\,\au$ to the West.  While both
are well-represented by \citet{pac86} curves (blue), they have substantially
different maximum magnifications and times of maximum, whose differences
yield a measurement of the ``microlens parallax'' vector $\bpi_\e$.
The dashed portion of the {\it Spitzer}
curve extends the model to what {\it Spitzer} could have observed
if it were not prevented from doing so by its Sun-angle constraints.  Light
curves are aligned to the OGLE $I$-band scale (as is customary), even though
{\it Spitzer} observations are at $3.6\,\mu$m.  Lower panel shows residuals.  
}
\label{fig:lc}
\end{figure}

As expected \citep{refsdal66}, the fit yields four distinct minima,
which are listed in Table~\ref{tab:ulens_1}.  
The best fit is shown in Figure~\ref{fig:lc}.
The remaining three fits look almost identical and so are not
shown to avoid clutter.

We note that the degeneracy between the $\Delta u_{0,-,\pm}$ and
$\Delta u_{0,+\pm}$ is marginally broken, with the latter two
disfavored by $\Delta\chi^2=8$ and 17, respectively.  However, the two 
$\Delta u_{0,-,\pm}$ solutions are consistent with each other at $<1\,\sigma$.

In Table~\ref{tab:ulens_1}, 
we have fit with blending as a free parameter for both observatories.
The results show that for the preferred solutions, the best-fit
blending for OGLE is negative but consistent with zero at the $1\,\sigma$
level.  A low level of negative blending is permitted because the
baseline photometry is carried out against a mottled background of
unresolved turnoff stars, and the source can in principle land on a ``hole''
in this background.  However, plausible levels of negative blending
due to this effect are $F_S\sim -0.2$ (on a flux scale of $I=18$ corresponding
to one flux unit), which is an order of magnitude smaller than what is
observed.  The most likely explanation is that the blending is very
small or zero and has fluctuated below zero in the fit because of the
relatively large errors in this quantity,
which are typical for low-amplitude microlensing events.

\begin{figure}
\plotone{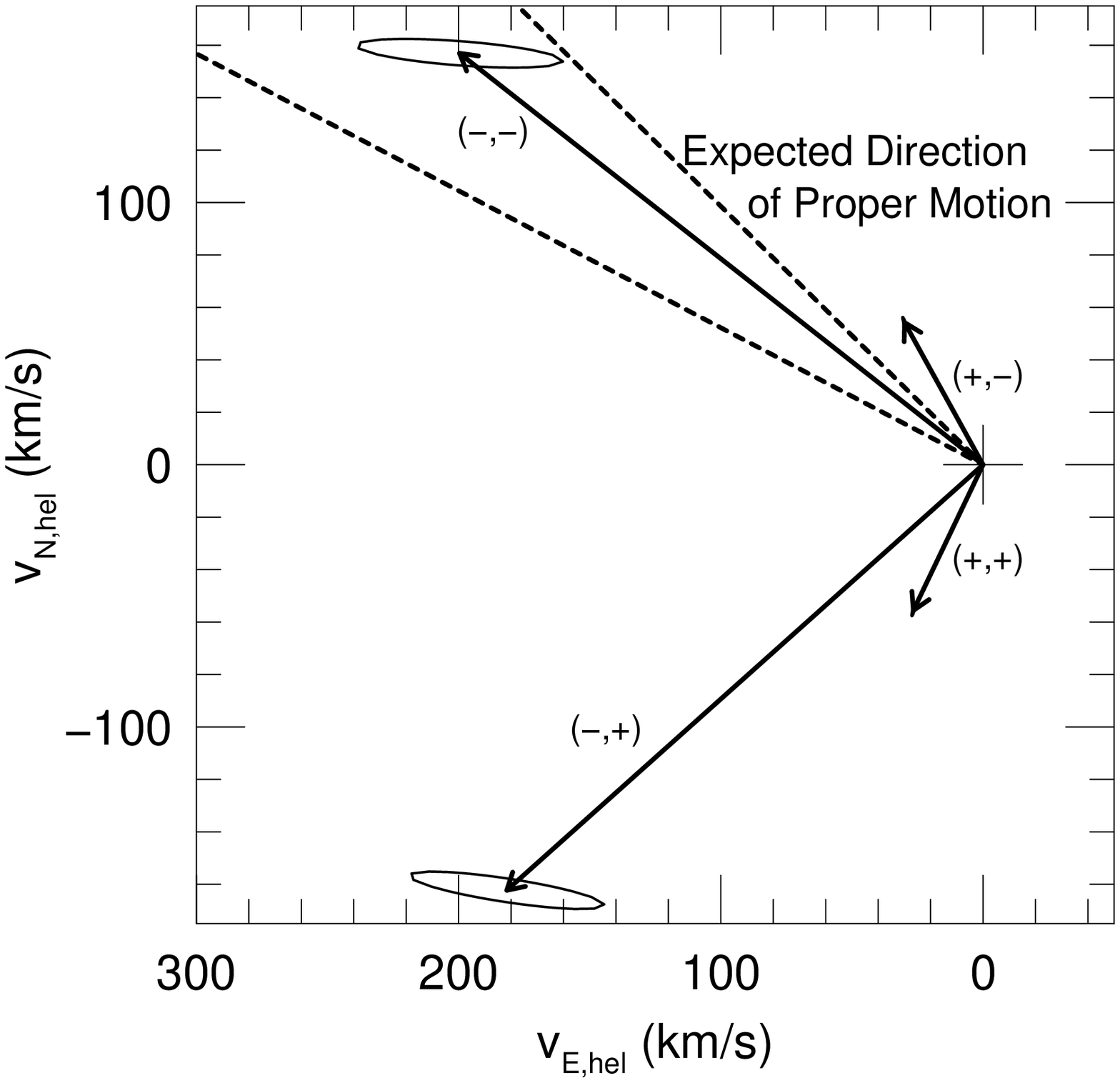}
\caption{Four-fold degeneracy in the heliocentric projected velocity 
$\tilde \bv_\hel =  \tilde \bv_\geo + \bv_{\oplus,\perp}$ where
$\tilde \bv_\geo = \bpi_{\e,\geo}\au/\pi_\e^2 t_\e$ and
$\bv_{\oplus,\perp}$ is the velocity of Earth projected on the sky
at the peak of the event.  Solutions are labeled $(\pm,\pm)$ by their
$\Delta u_0$ degeneracy.
Two smaller $\tilde v_\hel$ $(+,\pm)$
are disfavored by $\Delta\chi^2=8$ and 17.  Note that the error ellipses
for these are quite small and partly obscured by the ``arrow heads''.
The dashed curves show the $1\,\sigma$ error for the expected 
direction $\tilde \bv_\hel$ (same as $\bmu_\hel$) based on the measured
proper motion of the source and the assumption that the lens is in the Galactic 
Disk.  This proper motion measurement decisively breaks the degeneracy.}
\label{fig:projvel}
\end{figure}

In addition to the parameter values, in Table~\ref{tab:ulens_1} we also list
the heliocentric
projected velocity $\tilde\bv_\hel$,
\begin{equation}
\tilde\bv_\hel = \tilde\bv_\geo + \bv_{\oplus,\perp};
\qquad
\tilde\bv_\geo = {\bpi_{\e,\geo}\over\pi_\e^2}\,{\au\over t_{\e,\geo}},
\label{eqn:vhel}
\end{equation}
where $\bv_{\oplus,\perp}{\rm (N,E)}\simeq(-0.5,28.9)\,\kms$ is the
projected velocity of Earth  at the peak of the event and
where we have explicitly noted that $\bpi_\e$ and $t_\e$ are evaluated
in the geocentric frame (as in Table~\ref{tab:ulens_1}).
Figure~\ref{fig:projvel} shows the projected velocities and $1\,\sigma$ error
ellipses for each of the four solutions.

We also show in Table~\ref{tab:ulens_2} 
the parameter values and errors under the
assumption that $F_B=0$.  As expected from the fact that $F_B$ was
consistent with zero, the central values hardly change after application
of this restriction.  Note also that while the errors in $u_0$, $t_\e$,
and $\pi_{\e,E}$ (all of which are strongly correlated with $F_B$)
shrink dramatically under this assumption, the errors in $\tilde\bv_\hel$
hardly change.  This is because the East component of $\tilde\bv_\hel$
(the one that is heavily correlated with $t_\e$) is directly determined
from $\Delta t_0$ together with the 
physical separation between {\it Spitzer} and
Earth at the times of the respective peaks, both of which are direct
empirical quantities, which do not depend on the fitted Einstein timescale
$t_\e$.

{\section{Interpretation}
\label{sec:interp}}

Here we illustrate the power of measuring $\pi_\e$ for estimating
the mass and distance, even when $\theta_\e$ is not measured or constrained
by considering the specific example of OGLE-2014-BLG-0939.

The $\Delta u_{0,-\pm}$ solutions are significantly favored by $\chi^2$
so we consider these first.  The solutions are nearly identical
except that $u_0$ and $\pi_{\e,N}$ reverse sign.  This is expected
under the ``ecliptic degeneracy'' \citep{ob09020}, which is particularly
strong in the present case because the source lies only $2^\circ$ from
the ecliptic.  

The magnitude of $\tilde v_\hel\sim 250\,\kms$ strongly favors a Galactic disk
lens at intermediate distances, an inference that follows from the
relation between  $\tilde\bv$ and $\bmu$
\begin{equation}
\bmu = {\tilde \bv\over \au}\pi_\rel.
\label{eqn:vtildemu}
\end{equation}
If the lens were in the Bulge ($\pi_\rel\la 0.02$), then this would
imply relative proper motion $\mu_\hel = 1.05\,\masyr(\pi_\rel/0.02)$.
This compares to typical Bulge lens-source proper motions $\mu\sim 4\,\masyr$.
Since the probability of an event scales $\propto \mu^2$, Bulge lenses
are strongly disfavored but not ruled out by this argument.  On the other hand,
for nearby lenses $(\pi_\rel \simeq \pi_L)$, the projected velocity
$\tilde \bv$ is nearly equal to the space velocity of the lens in the
frame of the Sun, $\bv_\perp$.  Since there are very few stars moving at these
speeds, this hypothesis is also disfavored.

At intermediate distances, we would expect that the lens-source motion
would be dominated by the fact that both the observer and lens partake
in the same flat rotation curve.  Thus, apart from the peculiar motion
of the Sun and the lens (and random ``noise'' introduced by the
proper motion of the source), we expect the lens to be moving in the
direction of Galactic rotation ($\sim 30^\circ$ East of North) at the
proper motion of SgrA*, $\mu_{\rm sgrA*} = 6.4\,\masyr$.  In fact, one
of these two solutions $(\Delta u_{0,-,-})$ does show motion similar
to this direction ($52^\circ$ East of North), making it the preferred
solution.

To make a more precise comparison between the expected and observed
heliocentric motions, we measure the proper motion of the ``source''
(actually the ``baseline object'' that is coincident with the source)
using four years of OGLE-IV data.  We find,
\begin{equation}
\bmu_{S,\hel}({\rm N,E}) = (-0.64\pm 0.45,-5.31\pm 0.45)\,\masyr.
\label{eqn:musource}
\end{equation}
In principle it is possible that this ``baseline object''
is a blend of two or more stars.  However, because the blending $F_B$
from the microlensing fit is consistent with zero and because the
surface density of stars that are bright enough to substantially
affect the proper motion measurement is small, we tentatively assume that the
proper motion of the microlensed source and this ``baseline object''
are the same.

Thus, ignoring the peculiar motion of the lens, we then expect
\begin{equation}
\bmu_{\rm exp,\hel} = \mu_{\rm sgrA*} \hat\bphi -\bmu_{S,\hel}
= (6.2\pm 0.5,8.5\pm 0.5)\,\masyr
\label{eqn:muexp}
\end{equation}
where $\hat\bphi$ is the direction of Galactic rotation.  The
direction of this proper motion $53.9^\circ\pm 2.7^\circ$ East of North.
We show immediately below that when account is taken of the dispersion
in lens peculiar motions, the error bar widens to 
\begin{equation}
\tan^{-1}{\mu_{\rm exp,\hel,E}\over \mu_{\rm exp,\hel,N}} =53.9^\circ\pm 8.5^\circ.  
\label{eqn:muhelang}
\end{equation}
This $1\,\sigma$ range of proper motions
is shown on Figure~\ref{fig:projvel}, which demonstrates that
$\bmu_{\rm exp,\hel}$ agrees extremely well with $\tilde \bv_\hel$ for
the $\Delta u_{0,-,-}$ solution, and disagrees with all the other
solutions.  Of the three other solutions, only $\Delta_{0,+,-}$ has
a direction of $\tilde\bv_\hel$ even remotely close to $\bmu_{\rm exp,\hel}$,
and this solution is disfavored by $\Delta\chi^2=17$ 
(see Tables~\ref{tab:ulens_1} and \ref{tab:ulens_2}, and 
Section~\ref{sec:dbi}, below).

Therefore, the degeneracy is decisively broken by the combination
of the measurement of the source proper motion $\bmu_S$ and the fact
that the value of $\tilde v_\hel$ strongly indicates that the lens
is in the Galactic disk.  This is a new form of degeneracy breaking
that was not previously anticipated.

We then apply Equation~(\ref{eqn:vtildemu}) to derive
\begin{equation}
\pi_\rel \sim \mu_{\rm exp,\hel}{\au\over\tilde v_\hel}= 0.20\,{\rm mas},
\qquad
M = {\pi_\rel\over\kappa \pi_\e^2} \sim 0.23\,M_\odot
\qquad (\Delta u_{0,-,-}).
\label{eqn:masdis}
\end{equation}
Note that by inserting $\mu_{\rm exp,\hel}$ into the first expression
in Equation~(\ref{eqn:masdis}), we are essentially applying the
method described in Section~\ref{sec:intro} 
(paragraph below Equation~(\ref{eqn:piedef})), except that
we are making a more precise estimate of $\mu_\hel$, which
is possible because the lens is already identified as being in the disk
and because we have a measurement of the source proper motion.

What is the precision of these estimates?  The error in the expected
proper-motion estimate along the direction of motion is about 5\%,
while the error in $\tilde \bv_\hel$ in this direction is about 8\%.
The direction of motion is inclined $\sim 22^\circ$ to the Galactic
plane.  Adopting dispersion of $18\,\kms$ perpendicular to and $33\,\kms$ 
parallel to the Galactic plane, 
we derive dispersions of $20\,\kms$ perpendicular
to and $30\,\kms$ parallel to the direction of motion.  These must
be multiplied by $\pi_L/\pi_\rel \sim 1.6$ to project them on the
observer plane, i.e., $32\,\kms$ and $48\,\kms$, respectively.  The former
was added in quadrature to the proper motion measurement error to
obtain the error bar in Equation~(\ref{eqn:muhelang}).
The latter contributes 19\% to the error in the 
comparison of amplitudes.  Combining these in quadrature yields 
$\pi_\rel=0.20\pm 0.04\,{\rm mas}$, or $D_L = 3.1\pm 0.4\,\kpc$.
The error in $M$ can be estimated from 
$4 G M/c^2=\tilde v_\hel\mu_\hel t_{\e,\hel}^2$.  The first two terms
have the same fractional error as above (25\%), with only a very
small fraction of this contributed by the lightcurve.  Therefore
it is appropriate to treat the error in the last term (18\%) as
uncorrelated, which yields $M=0.23\pm 0.07\,M_\odot$.

{\section{Degeneracies and Degeneracy-Breaking Information From the Lightcurve}
\label{sec:degen2}}

As discussed in Section~\ref{sec:interp}, we have decisively broken the
four-fold degeneracy by measuring the source proper motion $\bmu_S$ and
taking advantage of the fact that the lens lies in the Galactic Disk,
which has well-organized motion.  However, it is also useful to ask
how well this degeneracy can be broken from lightcurve information alone
since, in general, source proper motion measurements can be very difficult
or impossible and not all lenses are in the Galactic Disk (or, more importantly,
can be localized as being in the Disk).  

{\subsection{Four-Fold Degeneracy}
\label{sec:4fold}}

\begin{figure}
\plotone{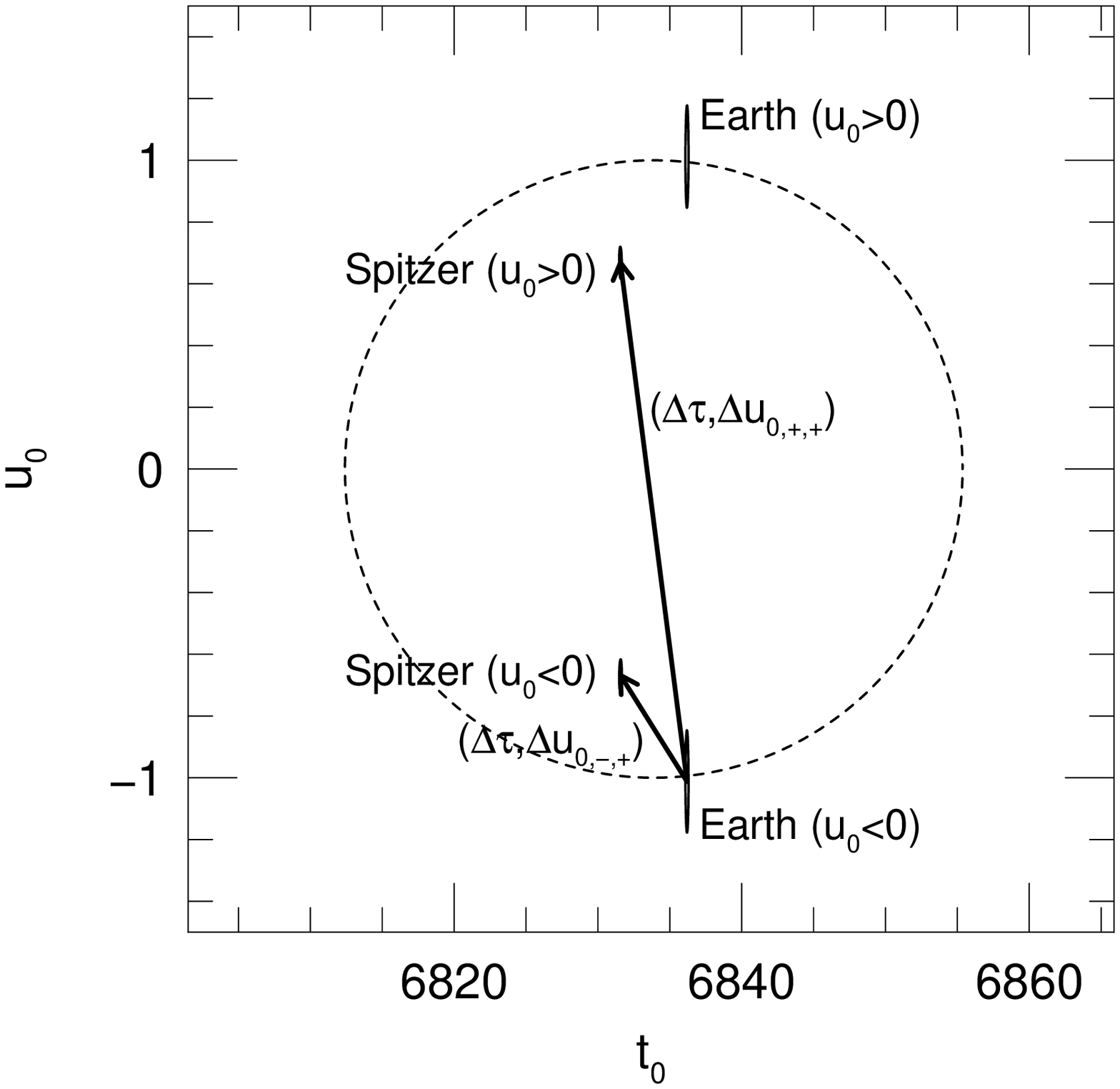}
\caption{Origin of \citet{refsdal66} four-fold degeneracy.  Lightcurves
from Figure~\ref{fig:lc} unambiguously determine peak times $t_0$ (abscissa)
as seen from Earth and {\it Spitzer} but only specify $u_0$ (ordinate)
up to a sign.  Hence, there are four ways to ``connect'' the Earth and
{\it Spitzer} measurements, which in each case is identified with
the Earth-{\it Spitzer} projected separation ${\bf D}_\perp$ to determine
the microlens parallax vector $\bpi_\e$ according to Equation~(\ref{eqn:sbpar}).
Dashed circle represents the Einstein radius, which brings the two axes to the
same system by scaling the abscissa by the Einstein timescale $t_\e$.
For each possible solution, the connecting line segment divided by $D_\perp$
is equal to $\pi_\e/\au$.  Two such line segments are shown explicitly, 
with $\Delta\tau\equiv\Delta t_0/t_\e$.
Hence there is a four-fold degeneracy in the
direction of $\bpi_\e$ but only a two-fold degeneracy in its magnitude.
Error ellipses for each solution generate much smaller errors, which
become important only if the discrete degeneracy is broken.
}
\label{fig:4fld}
\end{figure}

Figure~\ref{fig:4fld} (modeled on Figure~1 of \citealt{gould94})
gives a schematic view of the major sources of information
that go into the parallax measurement and thus into the origins of
the discrete and continuous degeneracies.  The larger ``ellipses''
(which are so flattened that they look like line segments)
represent the measurements of $t_0$ and $u_0$ based on a fit
to OGLE data assuming rectilinear lens-source relative motion, i.e.,
according to Equation~(\ref{eqn:pac}).  Properly speaking, we should
plot results of a similar fit for the {\it Spitzer} lightcurve.  However,
because there are no wing or baseline data from {\it Spitzer}, such
a fit would be extremely poorly constrained. Instead, we therefore
plot the results of a fit with the {\it Spitzer} timescale fixed
at the best-fit OGLE value.  This is legitimate because in the
combined fit to the data, the {\it Spitzer} timescale is very tightly
constrained by the OGLE timescale, although the constraint is
slightly offset from equality due to Earth-{\it Spitzer} relative
motion and lens-source relative parallax.
The parameters of these single-observatory (OGLE or {\it Spitzer})
fits are listed in Table~\ref{tab:ulens_3}.

Within the framework of this diagram, any line segment can be drawn
from an OGLE ellipse to a Spitzer ellipse.  The length of this
line segment relative to the radius of the Einstein ring (large circle)
corresponds to the length of Earth-{\it Spitzer} projected separation
${\bf D}_\perp$ relative to the projected Einstein radius 
$\tilde r_\e\equiv \au/\pi_\e$.  That is, 
$\pi_\e = (\Delta t_0^2/t_\e^2 +\Delta u_0^2)^{1/2}\au/D_\perp$.  Similarly,
the direction of the line segment gives the direction of $\bpi_\e$
 according to Equation~(\ref{eqn:sbpar}).  Four classes
of line segments can be drawn, corresponding to the four-fold degeneracy.
In addition, within each class, there is some freedom 
(primarily in the vertical direction to place the line segments within the 
two error ellipses.  Not represented in this diagram is the fact that
Einstein timescale $t_\e$ also has an error bar, so that while $\Delta t_0$
is extremely well determined,  the fractional error in $\Delta t_0/t_\e$
(the quantity going into $\bpi_\e$) is basically the same as the fractional
error in $t_\e$.  Nevertheless, since such errors are usually modest
($\sim 10\%$ in the present case), the fractional errors in $\Delta u_0$
are likely to be larger, particularly for $\Delta u_{0,-\pm}$.  Thus,
$\Delta u_0$ direction is generically most problematic both because
it suffers from a four-fold discrete degeneracy and because each of the
four local error ellipses are elongated in the $\Delta u_0$ direction.

{\subsection{Degeneracy-Breaking Information}
\label{sec:dbi}}

There are two striking differences between the full solution
presented in Table~\ref{tab:ulens_1} and the schematic solution
presented in Figure~\ref{fig:4fld} and Table~\ref{tab:ulens_3}.  First, many of
the geocentric parameters in Table~\ref{tab:ulens_1} are better constrained 
than the OGLE-only parameters in Table~\ref{tab:ulens_3}.  This includes
$u_0$, $t_\e$, and especially $F_S$ and $F_B$.  These parameters
are  strongly correlated, so it is not surprising that if the
errors in one are improved, then all will be improved.  Nevertheless,
this result is puzzling because the OGLE and {\it Spitzer} data
appear to couple only through $t_\e$, and we have already noted that
the {\it Spitzer} data by themselves contain virtually no information
about $t_\e$.

Second, from the standpoint of the simple \citet{pac86} fits that
are tabulated in Table~\ref{tab:ulens_3} and whose differences are
displayed in Figure~\ref{fig:4fld}, 
the $u_{0,+,\pm}$ and $u_{0,-,\pm}$
solutions appear equally good.  That is, the $\pm u_0$ solutions
shown at the top and bottom of Figure~\ref{fig:4fld} produce {\it exactly}
the same lightcurve in Equation~(\ref{eqn:pac}), so there cannot
be any $\chi^2$ difference between one combination of these and another.
However, according to Table~\ref{tab:ulens_1}, 
the $u_{0,-,\pm}$ solutions are clearly
preferred.

What is the source of additional information that reduces the parameter
errors and discriminates between the four discrete solutions when
the two lightcurves are fit simultaneously relative to when they
are fit separately?

The answer cannot be either of the two previous suggestions that
were summarized in Sections~\ref{sec:deltate} and \ref{sec:1Ds}.  
As just noted, the
{\it Spitzer} data by themselves contain essentially no timescale
information, so $\Delta t_\e$ cannot be measured and hence cannot
be used to discriminate among solutions with different $\Delta u_0$.
In addition, because the field lies extremely close to the ecliptic,
$\Delta t_\e$ would give information about the parallax in a direction
that is very nearly parallel to $\Delta t_0$ (i.e., ${\bf D}_\perp$
axis).  And, for the same reason (as already noted by \citealt{gould99}),
the instantaneous 
Earth (or satellite) acceleration is almost perfectly aligned with
${\bf D}_\perp$, which implies that the ``1-D parallax'' due to this
instantaneous acceleration \citep{gmb94} provides almost no 
information about $\Delta u_0$.

For the second effect (discrimination between discrete minima)
the answer turns out to be a previously unrecognized source of 
degeneracy-breaking information.  The OGLE data, by themselves,
give an extremely crude 2-D parallax measurement (due to changing 
acceleration of Earth over the course of the event), 
so crude that it would not normally be considered of any use,
and indeed by itself would 
not be of use in the present case.  However, if we fix 
$\pi_{\e,E}=0.24$ (the preferred value for the $\Delta u_{0,-\pm}$ solutions),
the OGLE data by themselves yield $\pi_{\e,N}(u_0>0) = 0.85\pm 0.95$
and $\pi_{\e,N}(u_0<0) = -0.55\pm 0.54$, which are consistent with
the fitted values from the full fit ($-0.25$ and $+0.22$) at the 
1.2 and 1.4 sigma.  However, when fixed to $\pi_{\e,E}=-0.06$ 
(the preferred value for the $\Delta u_{0,+\pm}$ solutions), the OGLE
data by themselves yield
$\pi_{\e,N}(u_0>0) = 1.15\pm 0.91$
and $\pi_{\e,N}(u_0<0) = -0.76\pm 0.54$, which are in conflict with
the full-solution values at $2.7\,\sigma$ and $3.9\,\sigma$, respectively.  
These values
explain both the quantitative preference for the $\Delta u_{0,-,\pm}$
solutions and also why $\Delta u_{0,+,-}$ is substantially
more disfavored than $\Delta u_{0,+,+}$.

However, this ``hidden information'' at most partly explains the
first effect.  Imposing a mathematical constraint on $\pi_{\e,E}$ (to reflect
the physical constraint on $\Delta t_0$ coming from the combination
of data from Earth and {\it Spitzer}) does drive down the errors
in $(u_0,t_\e,F_S,F_B)$ relative to no constraint, but the errors in
these quantities are still larger than those in Table~\ref{tab:ulens_3}, 
which assume
$\bpi_\e=0$.  Moreover, the errors in the OGLE-only $\pi_{\e,N}$ measurement
are an order of magnitude larger than the (local solution) errors
in $\pi_{\e,N}$ from the joint fit.  Thus, they are useful only for
discriminating between widely differing $\pi_{\e,N}$ solutions but not
for the modest tightening of individual solutions.  Hence, the source of 
this aspect of the improvement remains unknown.

We do note, however, that the {\it relative} improvement in flux errors
compared to $u_0$ errors is well understood.  The peak flux $F_{\rm peak}$
and the baseline flux $F_{\rm base} = F_S + F_B$ are both extremely
robust parameters.  Hence, so is their difference:
\begin{equation}
F_{\rm peak} - F_{\rm base} =F_S\biggl({u_0^2 + 2\over\sqrt{u_0^4 + 4u_0^2}}-1\biggr).
\label{eqn:peak}
\end{equation}
Treating the left-hand side of this equation as a constant and differentiating
yields,
\begin{equation}
{\delta\ln u_0\over\delta\ln F_s} =
\biggl({u_0^2 + 2\over\sqrt{u_0^4 + 4u_0^2}}-1\biggr){u_0(u_0^2+4)^{3/2}\over 8}
\rightarrow 0.48
\label{eqn:diff}
\end{equation}
where the evaluation is for $u_0=1$.  Thus we expect that the
fractional improvement in $F_S$ will be about twice as great as that
in $u_0$.

Finally, we note that, overall, it is far more important to break
the discrete degeneracy than to tighten the errors on individual solutions,
so the understanding of the former that has been achieved is by
the same token more important than the remaining uncertainty about the latter.

{\subsection{Degeneracy Breaking From $(I-[3.6])_S$ Color}
\label{eqn:extra}}

As discussed in Section~\ref{sec:photo}, it may in principle be possible to
break the four-fold degeneracy by using external information to
determine the ``color'' (log of the ratio of source fluxes)
between bands used for observations from Earth and the satellite.
In our case, this would be the $(I - [3.6])_S$ color.  The usual
way to determine the color of a microlensed source is regression.
That is if, for example, a series of $V$ and $I$ flux measurements are taken
at nearly the same time, $F_V(t_i) = F_{S,V} A(t_i) + F_{B,V}$ and
$F_I(t_i) = F_{S,I} A(t_i) + F_{B,I}$, then without even having a model
to tell one the magnifications $A(t_i)$, one can write
$F_V(t_i) = a F_I(t_i) + b$, yielding $(V-I)_S=-2.5\log(a)+const$.  This
also implies that any model of the lightcurve must yield very similar
$(V-I)_S$ colors, assuming that there are substantial contemporaneous
magnified data in these two bands.

This logic breaks down for parallax observations because one
does not know a priori that the magnifications are the same
for contemporaneous observations.  Indeed, it is only if these
magnification differ that one can measure the parallax.  Thus,
different solutions may have different colors.  Indeed, 
Tables~\ref{tab:ulens_1} and \ref{tab:ulens_2}
show that the four solutions have substantially different instrumental
$(I - [3.6])_S$ colors, which range from $(I-[3.6])_S=-1.17$ to $-1.43$.
Immediately below, we briefly describe how we use the method of 
\citet{mb11293} to measure the source color to be $(I-[3.6])_S=-1.216\pm 0.044$.
However, including this measurement into the fits does not significantly
alter the $\chi^2$ differences among the four solutions.  The
reason appears to be that the color errors shown in Table~\ref{tab:ulens_1} 
are of
the same order as the color differences between solutions, so that
the solutions can accommodate constraints on the color within this
range without significantly changing $\chi^2$.

The problem would appear to be that we lack {\it Spitzer} baseline data, 
which substantially degrades the determination the $(I-[3.6])_S$ color.
For instance, if we put in an artificial baseline measurement
with a precision of 0.005 mag
(which could, e.g., be acquired in future {\it Spitzer} seasons),
we find that the color error from the fits is reduced by a factor $\sim 3$
from $\sim 0.20$ mag to $\sim 0.07$ mag.  However, including both
this artificial baseline measurement and our actual color measurement
only increases the $\chi^2$ difference between solutions from 8 to 10,
despite the fact that both our real color measurement and our artificial
baseline measurement agree perfectly with the preferred solution,
while they do not agree with the alternate solutions.

We conclude that, at least in this case, a fairly accurate $(I-[3.6])_S$
source color measurement is not of substantial value in distinguishing
between solutions.

For completeness, we outline our method of measuring the $(I-[3.6])_S$
source color, which is a variant of the method used by \citet{mb11293}.
In our first attempt, we constructed a $(I-[3.6])$ vs. $(V-I)$ 
color-color diagram by matching field stars in OGLE $(V/I)$
and {\it Spitzer} ([3.6]) photometry, using the same instrumental
system that was used for the lightcurve photometry.  We then
measured the $(V-I)_S$ source color from regression (as described above),
with an error of 0.026 mag.  However, because of the steep slope
and significant scatter in the color-color diagram, we found this approach
to be unsatisfactory.

Therefore, we used $H$-band data of the event taken with the ANDICAM
camera on the 1.3m CTIO-SMARTS telescope, combined with OGLE $I$-band
data to measure $(I-H)_S$, and so derived $(V-H)_S$, which has
a factor two longer wavelength baseline than $(V-I)$, i.e., a factor $\sim 3$
compared to a factor $\sim 1.5$.  Of course, these added steps led to
larger errors in the $(V-H)_S$ color (0.044 mag), but the color-color
diagram had a substantially shallower slope and also less scatter.
We note that for future events, a more precise $(I-[3.6])_S$ source color could
be obtained by an increased number of $V$ and $H$ band observations.

{\section{Future Mass Measurement}
\label{eqn:mass}}

As we have emphasized, the ensemble of single-lens parallax measurements 
can be used to infer the mass function of stars (and other objects)
in the field without any additional data.  In the present case, we
have shown that the four-fold degeneracy
is broken.  Whether broken, partially broken, or unbroken,
the ensemble of measurements can be tested against various trial
mass functions using a likelihood estimator.

However, here we point out that essentially all such parallax measurements
can be turned into individual mass (and distance and transverse velocity)
measurements by direct imaging of the lens.  We use OGLE-2014-BLG-0939
as a concrete example.

Figure~\ref{fig:projvel} 
shows the measured projected velocities (and $1\,\sigma$ error
ellipses) of the four solutions.  The essence of this new
method for measuring lens  masses
is simply to take a late-time high-resolution image (e.g., using adaptive
optics (AO)) of the source and lens after they have separated.  
 From the measured vector separation
$\Delta\btheta$ and the elapsed time $\Delta t$ (and for the moment making
the assumptions that the source and lens were coincident at the peak
of the event and that the image is taken at the same time of year
as the event), we can then derive the heliocentric proper motion,
\begin{equation}
\bmu_\hel = {\Delta\btheta\over \Delta t}.
\label{eqn:helio}
\end{equation}
Comparing the direction of this vector to the four, clearly distinct
directions of the solutions shown in Figure~\ref{fig:projvel} 
one can unambiguously
pick out the correct solution.  Then it is a simple matter to obtain
\begin{equation}
\pi_\rel = {\au\over\tilde v_\hel}\mu_\hel;
\qquad
M = {\pi_\rel\over \kappa \pi_\e^2}.
\label{eqn:mpirel}
\end{equation}

For example, the Giant Magellan Telescope (GMT) will have a FWHM in $J$ band
of $11\,{\rm mas}$.  
For typical events, the proper motion will be 3--7 $\masyr$,
and hence the source and lens will have separated by 2 FWHM in 3--7 yr.
In particular, by the time GMT is operational (perhaps 2024), it is very likely
that the lens and source of OGLE-2014-BLG-0939 will be separated enough to
make this measurement.

We now address various departures from our zeroth-order assumptions.  
First, the lens and source
are not coincident at peak but are separated by 
$\delta\theta = u_{0,\oplus}\theta_\e$.  However, since $\theta_\e\la 1\,$mas
while $\Delta\theta\ga 20\,$mas, this will not interfere with choosing
the correct solution from comparison to Figure~\ref{fig:projvel}.  
Then, once the correct
solution is known, the actual path of the lens relative to the source
will also be known, allowing $\bmu_\hel$ to be correctly estimated.

Second, the followup image may not be taken at the same time of year,
which would lead to parallax effects.
However, since $\pi_\rel<1\,$mas in essentially all cases, while
$\Delta\theta\ga 20\,$mas, this will again not interfere with choosing
the correct solution, and hence allowing for proper correction of
parallax effects using the known $\pi_\rel$.

Third, in a substantial minority of cases, the microlensing event will
be due to the less massive (and so less luminous) member of a binary
system.  When the AO image is taken, the brighter companion will be
mistaken for the lens, yielding an incorrect $\bmu_\hel$.  \citet{gould14}
discusses this problem in detail for the more difficult case that
a 1-D geocentric parallax has been measured (rather than the simpler 4-fold
discrete degeneracy under consideration here).  In the current context,
this will give rise to two types of discrepancy. First, the inferred
$\bmu_\hel$ will not agree with any of the directions of the four 
$\tilde \bv_\hel$ solutions.  Second, the inferred mass will not agree
with the photometric estimates based on the measured brightness and
inferred distance of the system.  In these cases, one can take a
second epoch of AO observations to measure $\bmu_\hel$ of the brighter
companion.  If the orbit is relatively tight 
(few AU, corresponding to $\la 1\,$mas) then the
apparent motion of the companion relative to the source
will be similar  to that of the lens, so
the original inferred proper motion will be correct, and it will
be realized that the lens was the fainter (unseen) companion.
If the orbit is more than a few AU, then the proper motion of the companion
between the first and second epochs will be very similar to the
proper motion of the lens during the event, so this companion proper
motion can just be used for $\bmu_\hel$.  In this case, one will be
able to derive the projected separation of the binary as well by
tracing the companion position back to the time of the microlensing
event when the lens had a known position relative to the source.
Note that the lack of binary signatures in the lightcurve will
exclude some range of binary companions.  In the case of high-magnification
events, this can include several decades of projected separation
(e.g., \citealt{mb11293B}), but even for more typical events the
exclusion range can be significant.

Fourth, in general, one needs to consider the impact of binary sources.
Well separated binary sources are not likely to be confused with the
lens because they are unlikely to lie in one of the four directions
allowed by the four-fold degeneracy.  In case of doubt, these can
be vetted by second-epoch observations in which they would show
common proper motion with the source.  Unresolved binary sources
might lead to displacement of the light centroid from the position
of the source.  This is relatively unlikely simply because microlensing
events are heavily biased toward brighter sources, while flux ratios
for solar-mass binaries tend to be high.  However, it is also
possible to vet against this possibility by comparing the source
flux derived from the lightcurve (i.e., $F_S$) with the observed
flux in the high-resolution image, to determine whether there is
any unresolved light.  In sum, the possibility of contamination
of the astrometric measurements by binary sources must be investigated
on a case by case basis, but generally is not expected to be a major
problem.

Finally, dark lenses (free-floating planets, brown dwarfs, neutron
stars, black holes, and some white dwarfs) will obviously not appear
in followup AO images.  To understand this case, let us consider how
such a non-detection would be interpreted from AO observations taken
10 years after the peak of OGLE-2014-BLG-0939.  For definiteness, we
will assume that if the lens were at least 20 mas from the source it
would have been detected.  Recall that there are basically two
solutions, $\tilde v_{\hel,-\pm} = 250\,\kms$ and $\tilde
v_{\hel,+\pm} = 60\,\kms$, with corresponding $\pi_\e\sim 0.35$ and
1.35, respectively.

Non-detection implies either that the lens is dark or 
that it is moving $\mu_\hel<2\,\masyr$.
In the latter case, according to Equation~(\ref{eqn:mpirel}) the
lens would have
$(\pi_\rel,M)=(<0.04\,{\rm mas},<0.04\,M_\odot)$ or
$(\pi_\rel,M)=(<0.16\,{\rm mas},<0.01\,M_\odot)$.
Thus, if it were moving too slowly to be seen (under the glare of the source)
then it would also be dark (specifically because it was substellar).  
Of course, this would not by itself allow
one to estimate its mass: it could be dark because it is a brown dwarf
or because it is a massive black hole.  However, applying a likelihood
function to an ensemble of such objects with microlens parallax measurements
that are definitely {\it known} not to be luminous will enable substantially
more precise reconstruction of the mass function than if the entire
ensemble of detections must be considered.

To illustrate this for OGLE-2014-BLG-0939, the assumption of an $M=5\,M_\odot$
black hole would imply $\pi_\rel = \kappa M \pi_\e^2$, which yields
5 mas and 75 mas, for $\Delta u_{0,-,\pm}$ and $\Delta u_{0,+,\pm}$,
respectively.  While it would be
very exciting to have such a black hole passing within 200 pc (or 13 pc)
of the Sun, the prior probability of this is extremely low, and it would
be highly discounted by any reasonable likelihood function.

{\section{Conclusions}
\label{sec:conclude}}

The lightcurves of OGLE-2014-BLG-0939 as seen from Earth and {\it Spitzer}
differ dramatically, with substantially different maximum magnifications
and times of maximum.  As predicted by \citet{refsdal66}, this allows to
measure the microlens parallax vector $\bpi_\e$ and corresponding
projected velocity $\tilde \bv$ up to a four-fold degeneracy.  

In Section~\ref{sec:interp} we have developed a new way to break this
degeneracy.  First, we show that the 
{\it magnitude} of the projected velocity $\tilde
v_\hel\sim 250\,\kms$, by itself, strongly favors a disk lens. If the
lens is then assumed to be in the disk, our measurement of the source
proper motion leads to a prediction for both the magnitude and
direction of the lens-source relative proper motion $\bmu_\hel$.  The
{\it direction} of $\bmu_\hel$ is then found to agree closely with
that of the $\tilde \bv_\hel$ of one of four solutions and is clearly
inconsistent with all of the other three.  The {\it magnitude}
of $\bmu_\hel$ then yields an estimate 
$\pi_\rel = \au\mu_\hel/\tilde v_\hel = 0.20\pm0.04\,$mas and
$M=\pi_\rel/\kappa\pi_\e^2 = 0.23\pm 0.07\,M_\odot$.
This new method is very powerful, but can only be applied to the
minority of events that are amenable to source proper-motion measurements.

In Section~\ref{sec:degen2}, we have investigated three of the four
ideas for breaking this degeneracy based on photometric data alone
that have been developed
over the past 20 years, as discussed in
Sections~\ref{sec:deltate}--\ref{sec:1Ds}.  The fourth idea
(Section~\ref{sec:himag}) is not applicable to the present case.  We
find that the degeneracy in the magnitude of these vectors is
basically broken, but the less important degeneracy in direction
remains intact.  We find that the mechanism for this degeneracy
breaking was not previously anticipated.  

We note that the $\Delta u_{0.-.-}$ solution picked out by the 
proper motion argument (Section~\ref{sec:interp}) is favored over 
the two $\Delta u_{0.+.\pm}$ solutions by $\Delta\chi^2=8$ and 17.
While such $\chi^2$ differences would not be completely convincing
on their own, as confirmation of the already strong proper-motion
argument, they are compelling.  In particular, of the three solutions
whose directions of $\tilde \bv$ conflict with the proper motion
argument, only the $\Delta u_{0,+,-}$ solution is remotely near consistency,
and this is disfavored in the lightcurve fit by $\Delta\chi^2=17$.
See Table~\ref{tab:ulens_1} and Figure~\ref{fig:projvel}.

An ensemble of such microlens parallax
measurements, which are currently being made
under our ongoing {\it Spitzer} program, can measure the single-lens
mass function, including dark objects.  We show that this measurement
could be improved substantially by high-resolution imaging of the
luminous lenses using, for
example, the Giant Magellan Telescope, roughly 10 years after the
{\it Spitzer}-Earth parallax measurement.

\acknowledgments

Work by JCY, AG, and SC was supported by JPL grant 1500811.  Work by JCY was
performed under contract with the California Institute of Technology
(Caltech)/Jet Propulsion Laboratory (JPL) funded by NASA through the
Sagan Fellowship Program executed by the NASA Exoplanet Science
Institute.
The OGLE project has received funding from the European Research Council
under the European Community's Seventh Framework Programme
(FP7/2007-2013) / ERC grant agreement no. 246678 to AU.
AG and BSG. were supported by NSF grant AST 1103471 
AG, BSG, and RWP were supported by NASA grant NNX12AB99G.
This work is based in part on observations made with the Spitzer Space
Telescope, which is operated by the Jet Propulsion Laboratory,
California Institute of Technology under a contract with NASA.

\begin{table}                                                                  
\caption{\label{tab:ulens_1}                                                   
\sc  $\mu$lens Parameters (Free $F_B$)}                                        
\vskip 1em                                                                     
\begin{tabular}{@{\extracolsep{0pt}}llrrrr}                                    
\hline                                                                         
\hline                                                                         
Parameter & Unit & $u_{0,-,+}$ & $u_{0,-,-}$ & $u_{0,+,+}$ & $u_{0,+,-}$ \\    
\hline \hline                                                                  
$\chi^2/$dof          &                & 273.1& 273.7& 281.5& 290.2\\
                      &                & /  265& /  265& /  265& /  265\\
 \hline
$t_0 - 6800$          &day             & 36.22& 36.20& 36.06& 35.95\\
                      &                &  0.11&  0.11&  0.11&  0.11\\
 \hline
$u_0$                 &                &   0.922&  -0.913&   0.897&  -0.843\\
                      &                &   0.132&   0.129&   0.125&   0.110\\
 \hline
$t_{\rm E}$           &day             &22.87&22.99&22.91&23.87\\
                      &                & 2.14& 2.12& 2.10& 2.04\\
 \hline
$\pi_{\rm E,N}$       &                &  -0.248&   0.220&  -1.370&   1.325\\
                      &                &   0.072&   0.067&   0.172&   0.158\\
 \hline
$\pi_{\rm E,E}$       &                &   0.234&   0.238&  -0.060&   0.024\\
                      &                &   0.028&   0.030&   0.025&   0.018\\
 \hline
$\tilde v_{\rm hel,N}$&km/s            &-162.3& 156.9& -55.5&  54.2\\
                      &                &   7.2&   5.5&   2.2&   2.1\\
 \hline
$\tilde v_{\rm hel,E}$&km/s            & 181.6& 199.7&  26.6&  29.9\\
                      &                &  37.2&  39.5&   0.7&   0.8\\
 \hline
$F_{\rm S,OGLE}$      &                & 13.20& 12.95& 12.51& 11.09\\
                      &                &  3.77&  3.63&  3.42&  2.75\\
 \hline
$F_{\rm B,OGLE}$      &                & -2.19& -1.93& -1.49& -0.08\\
                      &                &  3.77&  3.62&  3.42&  2.75\\
 \hline
$F_{\rm S,Spitzer}$   &                &  4.31&  4.37&  3.32&  3.30\\
                      &                &  1.10&  1.12&  0.72&  0.69\\
 \hline
$F_{\rm B,Spitzer}$   &                & -0.08& -0.15&  0.96&  1.02\\
                      &                &  1.21&  1.22&  0.81&  0.79\\
 \hline
 \hline
\end{tabular}                                                                  
\end{table}                                                                    

\begin{table}                                                                  
\caption{\label{tab:ulens_2}                                                   
\sc  $\mu$lens Parameters ($F_{B,\rm OGLE}=0$)}                                
\vskip 1em                                                                     
\begin{tabular}{@{\extracolsep{0pt}}llrrrr}                                    
\hline                                                                         
\hline                                                                         
Parameter & Unit & $u_{0,-,+}$ & $u_{0,-,-}$ & $u_{0,+,+}$ & $u_{0,+,-}$ \\    
\hline \hline                                                                  
$\chi^2/$dof          &                & 273.6& 274.1& 281.8& 290.2\\
                      &                & /  266& /  266& /  266& /  266\\
 \hline
$t_0 - 6800$          &day             & 36.22& 36.20& 36.07& 35.95\\
                      &                &  0.11&  0.11&  0.10&  0.11\\
 \hline
$u_0$                 &                &   0.840&  -0.840&   0.840&  -0.840\\
                      &                &   0.002&   0.002&   0.002&   0.002\\
 \hline
$t_{\rm E}$           &day             &24.29&24.27&23.92&23.93\\
                      &                & 0.16& 0.16& 0.15& 0.15\\
 \hline
$\pi_{\rm E,N}$       &                &  -0.214&   0.192&  -1.292&   1.321\\
                      &                &   0.044&   0.043&   0.029&   0.029\\
 \hline
$\pi_{\rm E,E}$       &                &   0.217&   0.222&  -0.052&   0.024\\
                      &                &   0.006&   0.008&   0.018&   0.033\\
 \hline
$\tilde v_{\rm hel,N}$&km/s            &-164.9& 158.3& -56.4&  54.3\\
                      &                &   4.8&   4.7&   1.3&   1.3\\
 \hline
$\tilde v_{\rm hel,E}$&km/s            & 195.5& 212.4&  26.7&  29.9\\
                      &                &  34.2&  36.3&   0.7&   0.8\\
 \hline
$F_{\rm S,OGLE}$      &                & 11.01& 11.01& 11.01& 11.01\\
                      &                &  0.00&  0.00&  0.00&  0.02\\
 \hline
$F_{\rm B,OGLE}$      &                &  0.00&  0.00&  0.00&  0.00\\
                      &                &  0.00&  0.00&  0.00&  0.00\\
 \hline
$F_{\rm S,Spitzer}$   &                &  3.85&  3.93&  3.10&  3.29\\
                      &                &  0.68&  0.69&  0.47&  0.50\\
 \hline
$F_{\rm B,Spitzer}$   &                &  0.34&  0.25&  1.15&  1.04\\
                      &                &  0.87&  0.88&  0.64&  0.66\\
 \hline
 \hline
\end{tabular}                                                                  
\end{table}                                                                    

\begin{table}                                                                  
\caption{\label{tab:ulens_3}                                                   
\sc  Single-Observatory Parameters}                                            
\vskip 1em                                                                     
\begin{tabular}{@{\extracolsep{0pt}}llrr}                                      
\hline                                                                         
\hline                                                                         
Parameter & Unit & OGLE & Spitzer \\                                           
\hline \hline                                                                  
$\chi^2/$dof          &                & 242.9&  28.0\\
                      &                & /  238& /   26\\
 \hline
$t_0 - 6800$          &day             & 36.20& 31.57\\
                      &                &  0.11&  0.09\\
 \hline
$u_0$                 &                &   1.012&   0.668\\
                      &                &   0.166&   0.052\\
 \hline
$t_{\rm E}$           &day             &21.48&21.48\\
                      &                & 2.31& 0.00\\
 \hline
$F_{\rm S}$           &                & 15.99&  4.32\\
                      &                &  5.47&  0.81\\
 \hline
$F_{\rm B}$           &                & -4.98& -0.13\\
                      &                &  5.47&  1.00\\
 \hline
 \hline
\end{tabular}                                                                  
\end{table}                                                                    

\end{document}